\documentclass[reprint,twocolumn,aps,showpacs,amsmath,amssymb,prl,superscriptaddress]{revtex4-1}
\usepackage{color}
\usepackage{graphicx}
\usepackage{float}
\usepackage[utf8]{inputenc}

\begin{document}

\title{Signatures of enhanced superconducting phase coherence through MID-IR excitation in optimally doped Y-Bi2212}

\author{Francesca Giusti}%
\affiliation{Department of Physics, Universit\`{a} degli Studi di Trieste, 34127 Trieste, Italy}

\author{A. Marciniak}%
\affiliation{Department of Physics, Universit\`{a} degli Studi di Trieste, 34127 Trieste, Italy}

\author{F. Randi}%
\affiliation{Department of Physics, Universit\`{a} degli Studi di Trieste, 34127 Trieste, Italy}

\author{G. Sparapassi}%
\affiliation{Department of Physics, Universit\`{a} degli Studi di Trieste, 34127 Trieste, Italy}

\author{F. Boschini}%
\affiliation{Department of Physics and Astronomy, University of British Columbia, Vancouver, Canada}
\affiliation{Quantum Matter Institute, University of British Columbia, Vancouver, BC V6T 1Z4, Canada}

\author{H. Eisaki}%
\affiliation{Nanoelectronics Research Institute, National Institute of Advanced Industrial Science and Technology, Tsukuba, Ibaraki 305-8568, Japan}

\author{M. Greven}%
\affiliation{School of Physics and Astronomy, University of Minnesota, Minneapolis, MN 55455, USA}

\author{A. Damascelli}%
\affiliation{Department of Physics and Astronomy, University of British Columbia, Vancouver, Canada}
\affiliation{Quantum Matter Institute, University of British Columbia, Vancouver, BC V6T 1Z4, Canada}

\author{Adolfo Avella}%
\email[Corresponding author: ]{avella@physics.unisa.it}
\affiliation{Department of Physics, Universit\`{a} degli Studi di Salerno, 84084 Fisciano (SA), Italy}

\author{Daniele Fausti}%
\email[Corresponding author: ]{daniele.fausti@elettra.eu}
\affiliation{Department of Physics, Universit\`{a} degli Studi di Trieste, 34127 Trieste, Italy}
\affiliation{Elettra Sincrotrone Trieste S.C.p.A., 34127 Basovizza Trieste, Italy}

\begin{abstract}
Optimally doped cuprate superconductors are characterized by the presence of superconducting fluctuations in a relatively large temperature region above the critical transition temperature. We reveal here that the effect of thermal disorder, which decreases the condensate phase coherence at equilibrium, can be dynamically contrasted by photoexcitation with ultrashort mid-infrared pulses. In particular, our findings reveal that light pulses with photon energy comparable to the amplitude of the superconducting gap and polarized in plane along the copper-copper direction $\left[110\right]$ can dynamically enhance the optical response which is associated to the onset of superconductivity. We propose that this effect could be rationalized by an effective d-wave BCS model, which reveals that mid-infrared pulses result in a transient increase of the phase coherence.
\end{abstract}

\maketitle

Many of the ingredients required for superconductivity in cuprates survive well beyond the region of the phase diagram where the actual macroscopic superconducting phase resides. An example of this hindered superconductivity is represented by the behavior of underdoped cuprates just above the critical temperature ($T_c$), where some hints indicate that pairing occurs, but the presence of the superconducting phase is inhibited either by a competing charge order or by the local nature of the pair incoherence (global phase incoherence), blocking the formation of a mesoscopic superconducting state \cite{pseudogap_rev, pseudogap1, pseudogap2, pseudogap3, pseudogap4}. Signatures of an incipient superconductivity at temperatures larger than $T_c$ have been revealed also in optimally doped samples, where the superconducting fluctuations survive tens of Kelvin above the actual $T_c$ \cite{perfetti, nature, ivan, damascelli, kondo}. The relative fragility of the superconducting phase, together with the underlying presence of its ingredients on large portions of the phase diagram, enables the possibility of controlling superconductivity through ultra-short light pulses.\par
While there is ample evidence that photo-excitation with ultrashort high photon energy pulses melts the superconducting phase under some specific conditions \cite{perfetti, damascelli, lanzara, Giannetti, ivan}, it has been shown that the formation of a superconducting phase can be triggered by mid-infrared (MIR) excitations in regions of the phase diagram that are not superconducting at equilibrium \cite{Fausti, Kaiser, Hu, Casandruc, Mitrano}. The possibility of triggering the onset of quantum coherence through MIR excitations could open up new avenues to control quantum states of matter through light.\\

Here, we contribute to the solution of the puzzle of light induced superconductivity by revealing that the time domain response of optimally-doped Yttrium substituted Bi2212 to excitations with photon energies close to the superconducting gap ($2\left|\Delta\right| \approx 75$  meV) is highly anisotropic. While excitations with pump polarization along the $\left[100\right]$ direction (or copper-oxygen Cu-O axis) lead to a reduction of the superconducting gap, independent of the photon frequency, photo-excitations along the $\left[110\right]$ direction (or copper-copper Cu-Cu axis) seem to trigger an increase of phase coherence, which results in an enhancement of the dynamical response associated with the superconducting phase.\par
Yttrium substituted Bi2212, $Bi_2 Sr_2 Y_{0.08} Ca_{0.92} Cu_2 O_{8+\delta}$ (Y-Bi2212), at optimal doping presents a superconducting phase below $T_c=97$ K, a pseudogap phase between $T_c$ and $T^* \approx 135$ K and an unusual ``strange-metal'' phase for higher temperatures \cite{Cilento}.\par
A much debated aspect of superconductivity in cuprates is that the onset of the superconducting phase is followed by a change of spectral weight at an energy scale orders of magnitude larger than the superconducting gap. In particular, in Bi2212, the opening of a superconducting gap at about 35-40 meV is accompained by a spectral weight redistribution at frequency as high as several eV \cite{static1, static2, static3}. This is visible in time domain studies where, upon a sudden perturbation of the superconducting gap, the reflectivity in the visible range changes consistently \cite{demsar2, gedik, kaindl, kusar, giannetti2, demsar3, averitt}. Here we leverage this characteristic and work under the assumption that the spectral response in the visible-near infrared region is intimately related to the dynamics.
\begin{figure}[ht]
\includegraphics[width=\linewidth]{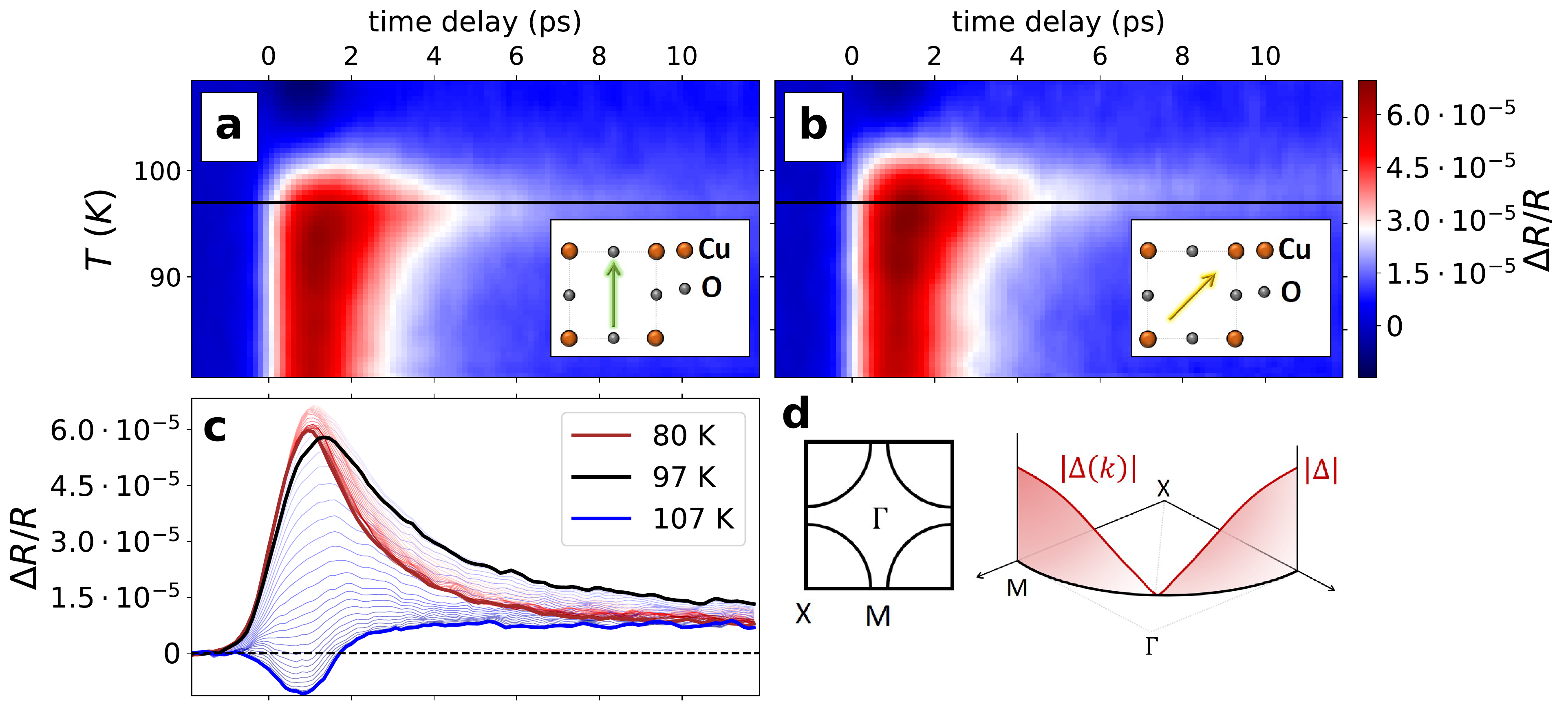}
\caption{\fontsize{9}{11} \textbf{MIR pump, optical probe measurements. }\selectfont a) and b) Reflectivity variation due to an impulsive excitation at time delay 0 by a MIR pulse ($h\nu \approx 70$ meV) as a function of the temperature of the sample (vertical axis).  The two maps differ for the polarization of the impinging pump, as highlighted by the two insets, representing the direction of the polarization (green and orange arrow) in the Cu-O plane. c) Time resolved signal at fixed temperatures for Bi2212: the brown line represents the characteristic superconductive signal, while the blue one refers to the pseudogap phase. The transition between the two phases is marked with the black line and is related to the divergence of the time decay. Light colored lines represent intermediate temperatures. d) Sketch of the first Brillouin zone and of the superconducting d-wave gap amplitude $\left|\Delta (\textbf{k})\right|$ in the reciprocal space. The black curved lines represent the Fermi surface.}
\label{MAP}
\end{figure}
%
We performed pump-probe measurements in which the sample has been excited with MIR ultra-short pulses ($h \nu \approx 70$ meV $\approx 2\Delta$, incident fluence $f=0.09$ $\mathrm{mJ \cdot cm^{-2}}$) and probed with near-infrared (NIR) ones ($h \nu \approx 1.55$ eV). Both pump and probe propagation directions are parallel to the c-axis (that is, perpendicular to the Cu-O layer) and their polarizations have been kept parallel in all measurements.\par
The intensity map shown in Figure \ref{MAP}a represents the relative variation of the reflectivity upon the pump excitation as a function of the time delay between the pump and the probe pulses (horizontal axis) and of the temperature of the sample (vertical axis). We focused our investigation in a temperature range across the superconducting-pseudogap transition (from 80 to 110 K). For $T<T_c$, the reflectivity increases for about 1 ps after the arrival of the pump (at 0 ps) and then it starts decreasing through an exponential decay (red line in Figure \ref{MAP}c). The characteristic time of the decay increases with temperature, it is maximum at $T=97$ K (green line) and drops for higher temperatures. The observed divergence of the time decay is a precise indicator of the superconducting-pseudgap phase transition \cite{divergence, Giannetti}. 
For higher temperatures, the sample enters the so-called pseudogap phase, whose time domain response is shown by the blue line in Figure \ref{MAP}c.\par

\begin{figure}[!ht]
\includegraphics[width=\linewidth]{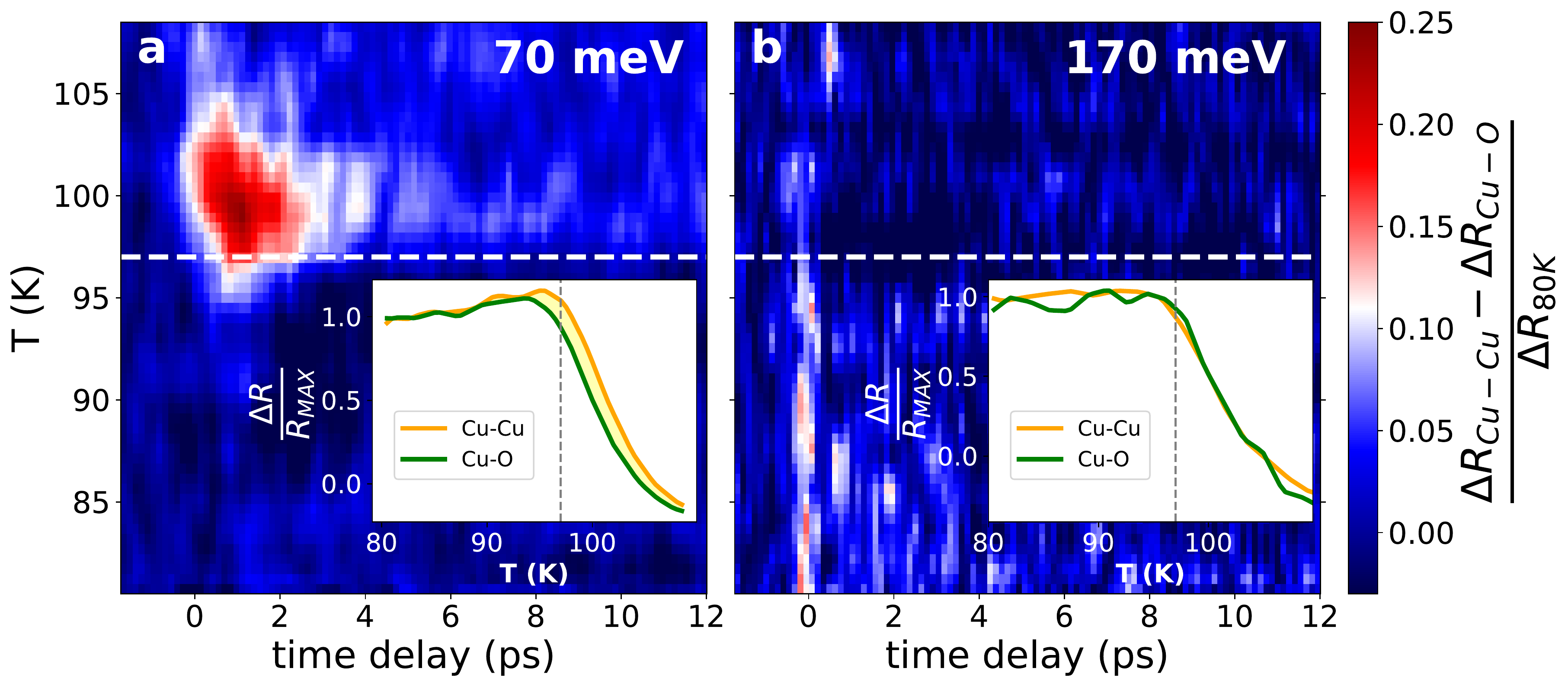}
\caption{\fontsize{9}{11}\textbf{Wavelength dependent anisotropy. }\selectfont Difference between the transient reflectivities due to Cu-Cu and Cu-O polarized pump in time (x-axis) and temperature (y-axis), induced by excitations with a) 70 meV and b) 170 meV pump photon energies. The dashed lines highlight the critical temperature $T_c=97$ K. The insets represent the response as a function of temperature at 1 ps time delay for Cu-Cu (orange line) and Cu-O (green line) polarized pump excitations for low and high pump photon energies (a and b, respectively). The gray dashed lines mark  $T_c$. Both in the maps and in the insets the values of the reflectivity have been normalized to the maximum value at 80 K ($\Delta R_{80K}$).}
\label{DIFF}
\end{figure}

The anisotropy of the gap of d-wave superconductors suggests a detailed analysis of the effects of excitations with different pump polarizations \cite{Devereaux}. Figure \ref{MAP}a and \ref{MAP}b show the measured transient reflectivity for two different pulse polarizations: along the Cu-O axis (Figure \ref{MAP}a) and the Cu-Cu direction (Figure \ref{MAP}b). We measured the transient reflectivity in both polarizations, using a pump photon energy around the characteristic energy of the system in the superconducting phase, that is $2|\Delta| \approx 75$ meV \cite{disegno_gap}. A convenient way to visualize the polarization dependence of our measurements is to subtract the two maps in Figure \ref{MAP}a and \ref{MAP}b, as displayed in the differential map of Figure \ref{DIFF}a. The difference map reveals a sizable signal around $T_c$, for time delays from 0 to 2 ps, corresponding to the maximum response in the superconducting phase (red region at about 1 ps in Figure \ref{DIFF}a).\par
The result is confirmed by the visual inspection of the temperature response at a fixed time delay (1 ps) for different polarizations of the pump, as shown in the inset of Figure \ref{DIFF}a.\par
We observe an increase of response associated with the onset of the superconductivity when the pump is polarized along the Cu-Cu direction both above and below $T_c = 97$ K. We stress that this is an anisotropic response strongly dependent on the photoexcitation wavelength:  in Figure 2b we display the differential map retrieved for higher pump photon energy ($h \nu \approx 170$ meV), which reveals no anisotropy at any temperature. This result has also been confirmed by measurements at high pump fluence (see Figure S4 of Supplemental Materials) demonstrating that the superconducting signal can be increased by a pump polarized along the Cu-Cu axis at long wavelength whereas higher photon energy excitations suppress this effect.\par
%
In order to draw a picture of the possible physical scenario emerging from the anisotropic response to low photon energy ultra-short pulses, we implemented a microscopic description based on a generalized BCS Hamiltonian allowing for a \textit{k}-dependent \textit{d}-wave gap (for further details see Supplementarl Materials). While it is well known that a simple BCS formalism, disregarding first and foremost the presence of electronic correlations, cannot explain the whole cuprate phenomenology, we will argue here that it accounts well for the non-equilibrium response of the low-energy superconducting gap, at least at a qualitative level. From the described Hamiltonian (see Supplemental Materials, Equation 1), it is possible, through density matrix formalism, to calculate the time evolution of several meaningful quantities (such as the superconducting gap amplitude $\left|\Delta\right|$).\par
\begin{figure}[ht]
\includegraphics[width=\linewidth]{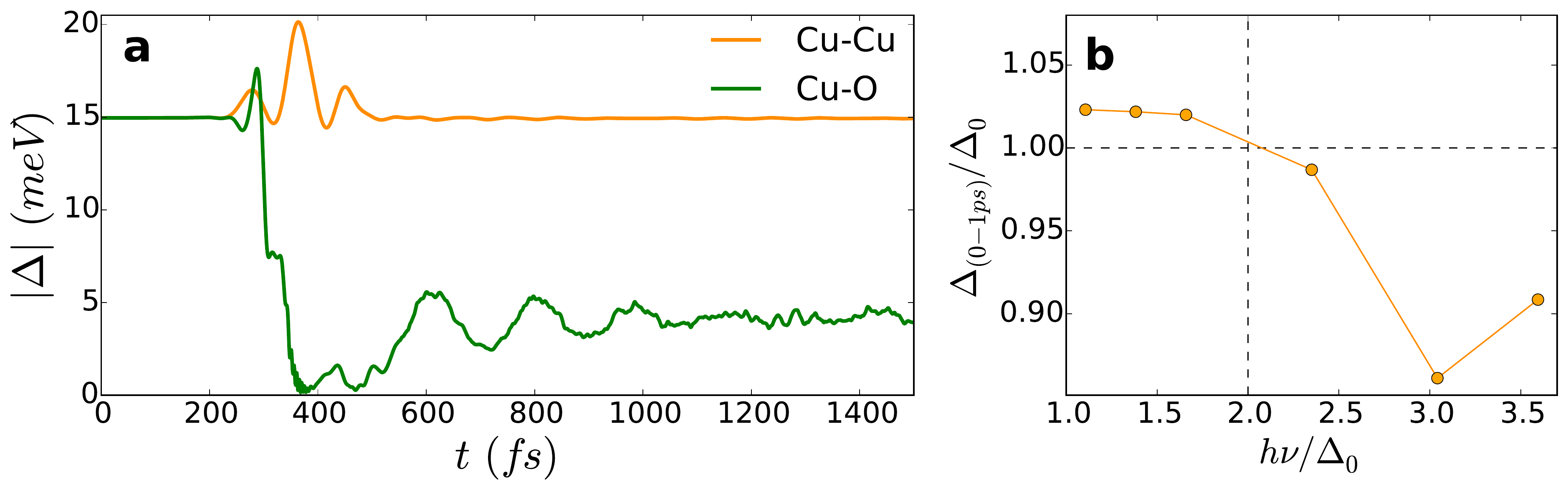}
\caption{\fontsize{9}{11}\textbf{d-wave BCS microscopic model. }\selectfont Results of the microscopic model: a) Time evolution of the modulus of the superconducting gap ($\left| \Delta \right|$) in both Cu-Cu (orange line) and Cu-O (green line) excitation case. The maximum of the pump electric field is reached at about 350 fs. b) Normalized integral of $|\Delta|$ in the time interval from 0 to 1 ps as a function of the pump photon energy ($\Delta_0=\left|\Delta\left(t=0\right)\right|$) for Cu-Cu polarized pump excitations.}
\label{DELTA}
\end{figure}

The model predicts different behaviors depending on the frequency and the polarization of the pump pulse. In particular, Figure \ref{DELTA}a shows that low photon energy excitations with polarization parallel to the Cu-Cu direction are predicted to drive an instantaneous enhancement of the superconducting gap, while a pump polarization rotated $45 ^{\circ}$ induces a dynamical quench of the gap. These results qualitatively rationalize the experimentally observed enhancement of the positive signal associated with the superconducting response, triggered by photoexcitation polarized along the Cu-Cu direction for pump photon energy $h\nu \approx 2 \left|\Delta\right|$. On the other hand, the collapse of the superconducting signal in the Cu-Cu polarization configuration is predicted for higher pump photon energies, as shown in Fig. \ref{DELTA}b, where we display the transient decreases of $|\Delta|$ as a function of the pump photon energy.\par
In order to grasp the physical picture that emerges from this microscopic model, we calculated the expectation value of the pair operator $\hat{\Psi} (\textbf{k})$ (see Supplemental Materials).\par
Figure \ref{MODEL} displays the modulus and the phase of the pair amplitude (in the first Brillouin zone) calculated in three different cases: at equilibrium and during a Cu-Cu and Cu-O low photon energy excitation ($h\nu \approx 2\Delta$). We observe that the value of $\left| \left\langle \hat{\Psi} (\textbf{k}) \right\rangle \right|$ around the Fermi surface is nearly unperturbed (and actually slightly enhanced) in both excitation schemes (Figure \ref{MODEL}b and \ref{MODEL}c), i.e., that pairing is still present in both cases. On the other hand, the phase reveals a strong anisotropic response (Figure \ref{MODEL}e and \ref{MODEL}f): the model shows that Cu-O polarized excitations drive intense phase fluctuations, which are responsible for the collapse of $|\Delta|$ shown in Figure \ref{DELTA}a. Cu-Cu polarized low-photon-energy excitations preserve instead phase coherence and enable an enhanced superconducting dynamical response.\par
\begin{figure}[ht]
\includegraphics[width=\linewidth]{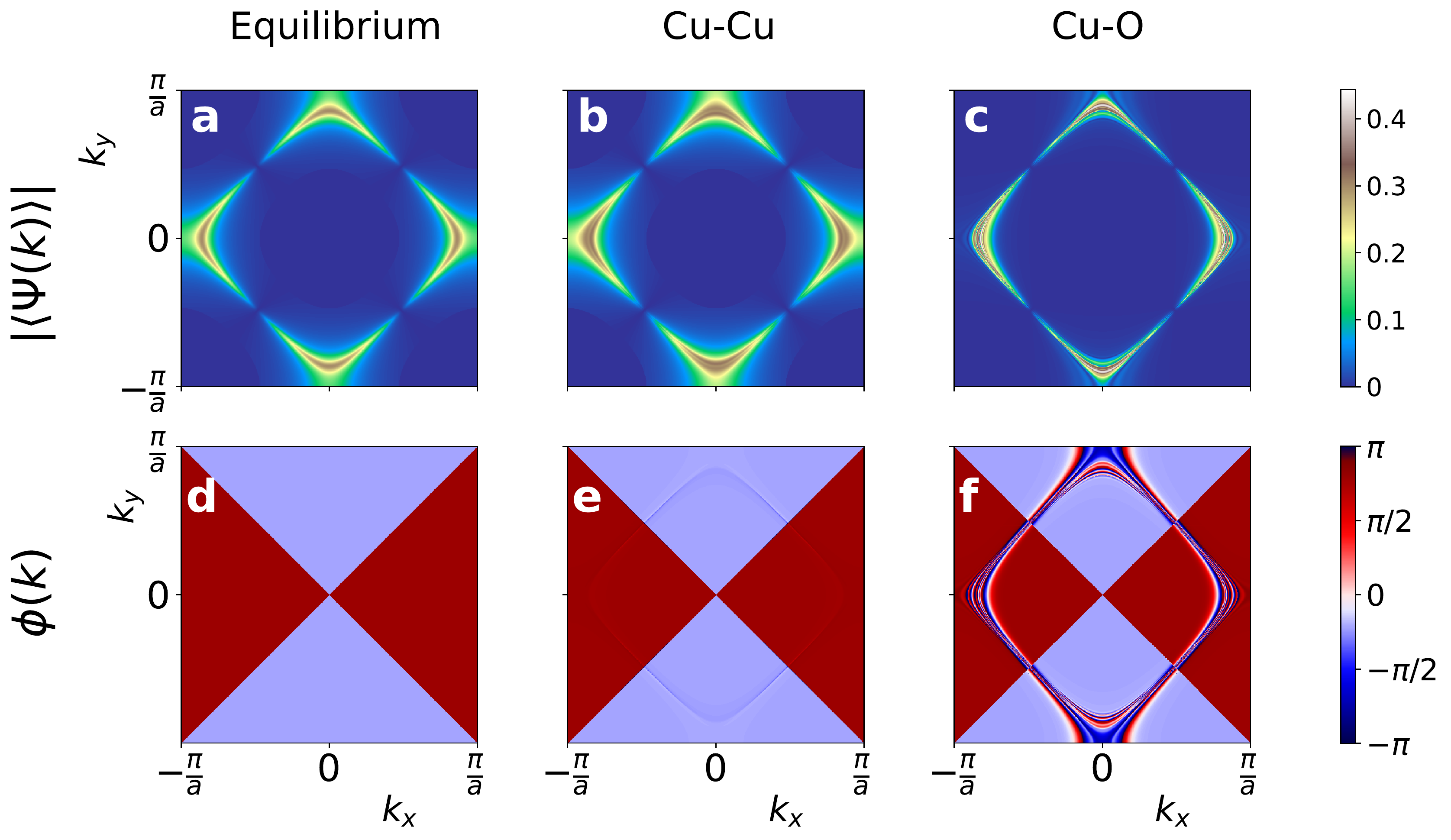}
\caption{\fontsize{9}{11}\textbf{Time dependent pair operator representation. }\selectfont a-c) Modulus of the expectation value of the pair operator $\hat{\Psi} (\textbf{k})$ in the reciprocal space in three different situations: a) no excitation (``Equilibrium''), b) during a pump excitation polarized along the Cu-Cu and c) Cu-O direction. Analogously, pictures from d to f show the phase $\phi (\textbf{k})$ of $\left\langle \hat{\Psi} (\textbf{k})\right\rangle$, in the same three conditions.}
\label{MODEL}
\end{figure}
%
We stress that the calculations were performed at a temperature $T<T_c$, where the amplitude of the superconducting gap $|\Delta|$  has a non-zero equilibrium value.
This is an intrinsic limitation of the microscopic BCS model, which does not allow superconducting pairing at temperatures higher than the critical value ($T_c$). The data instead report a well visible enhancement of the superconducting behavior at temperatures up to $\approx 10$ K above the equilibrium critical temperature $T_c$ (inset of Figure \ref{DIFF}a).\\
In order to extend this effective description to higher temperatures, we propose to run calculations from a modified equilibrium state, whose features are justified in the following, maintaining the BCS framework. Different from standard BCS superconductors, cuprates exhibit signatures of strong superconducting fluctuations at temperatures larger than $T_c$ \cite{nature}. In particular, in optimally-doped Bi2212, both equilibrium and time-domain techniques revealed superconducting fluctuations up to tens of Kelvin above the critical temperature \cite{ivan,damascelli,perfetti, kondo, nernst, diamag}. This anomalous feature is commonly taken to imply the presence of Cooper pairs losing phase coherence; i.e. while the mesoscopic coherence vanishes above the transition temperature, pairing remains, together with phase correlations, which are local in space and time \cite {nature}. Transport and magnetization studies suggest that the local correlations lead to a universal superconducting percolative regime above $T_c$ consistent with a percolation picture of the phase diagram \cite{greven4}; in particular, a local superconducting gap distribution was able to explain \cite{greven2} the presence of an effective average gap above $T_c$ in photoemission experiments \cite{reber} and the exponential tail exhibited by several observables. Here we observe qualitatively the same signatures of local correlations persisting up to $\sim 10$ K above $T_c$, a temperature range similar to previous studies \cite{greven1, greven2, greven3, reber}.\par
\begin{figure}[ht]
\includegraphics[width=\linewidth]{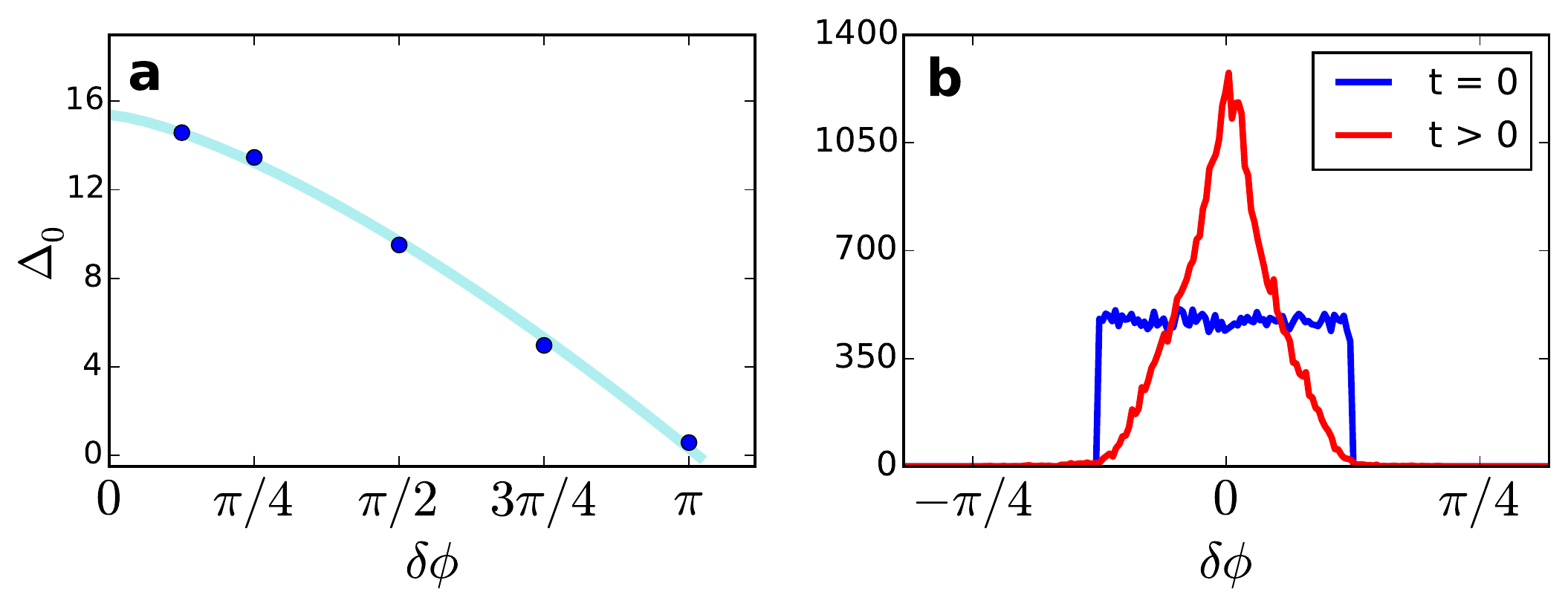}
\caption{\fontsize{9}{11}\textbf{Extension of the d-wave BCS model to larger temperatures. }\selectfont a) Values of the superconducting gap at equilibrium ($\Delta_0 = \left| \Delta \right|(t=0)$) as a function of the amplitude of the phase noise. b) Distribution of the phase values at equilibrium (blue line) and during the excitation polarized along the Cu-Cu direction (red line) for an initial noise amplitude of $\delta \phi = \pi/8$.}
\label{NOISE}
\end{figure}

In order to explore these effects within the generalized BCS model, we employ the following simple procedure. We proposed a new equilibrium state artificially built by adding a random noise to the phase of the original state pair amplitude $\phi (\textbf{k})$, while retaining its modulus $\left(\left| \left\langle \hat{\Psi}(\textbf{k}) \right\rangle \right| \right)$. The phase noise introduced in the model leads to a reduction of the gap, as shown in Figure \ref{NOISE}a, in which the dependence of the gap amplitude on the maximum value for the phase fluctuations $\left(\delta \phi\right)$ is plotted. Calculations of the dynamic response, starting from this inhomogeneous (in momentum space) equilibrium state, reveal that Cu-Cu low-photon-energy excitations induce a significant enhancement of phase coherence (and negligible variation in the amplitude) of the pair operator, as highlighted in Figure \ref{NOISE}b, which depicts a histogram of the phase distribution before (blue line) and during (red line) the photo-excitation (350 fs), for an initial fluctuation of $\delta \phi = \pi/8$ (value chosen for sake of clarity). The plot reveals that the phase distribution, which becomes narrower after the excitation, leads to an enhanced superconducting response. We argue that this scenario rationalizes the enhancement of an out-of-equilibrium superconducting behavior above $T_c$, which could therefore be associated with a light-driven boost of phase coherence.\par
The scenario that emerges from our pump-probe experiments reveals the capability to enhance the  transient response associated with superconducting fluctuations in cuprates by means of photo-excitations with low-energy photons polarized in the Cu-Cu direction, which is completely suppressed by a pump excitation with polarization parallel the Cu-O axis. The effective \textit{d}-wave BCS description of the interaction of the superconductor and pulsed electromagnetic radiation is in qualitative agreement with the experimental results. Moreover it allows us to ascribe the observed dynamical increase of the superconductive response to a light-driven enhancement of phase coherence below and above $T_c$, where thermodynamic constraints make the superconducting equilibrium state unattainable. The supposed field-driven increase of phase coherence highlights the possibility of driving the onsets of quantum coherence in complex oxides.\par
%

This research was undertaken thanks in part to funding from the Max Planck UBC Centre for Quantum Materials and the Canada First Research Excellence Fund, Quantum Materials and Future Technologies Program. The work at UBC was supported by the Killam, Alfred P. Sloan, and Natural Sciences and Engineering Research Council of Canada (NSERC) Steacie Memorial Fellowships (A.D.), the Alexander von Humboldt Fellowship (A.D.), the Canada Research Chairs Program (A.D.), NSERC, Canada Foundation for Innovation (CFI) and CIFAR Quantum Materials Program. The work at the University of Minnesota was funded by the Department of Energy through the University of Minnesota Center for Quantum Materials under DE-SC-0016371. D.F acknowledges the support by the European Commission through the ERCStG2015, INCEPT, Grant No. 677488.\\

\end{document}